\documentclass[10pt,aps,prd,twocolumn,notitlepage,nofootinbib,floatfix,superscriptaddress]{revtex4-1}

\RequirePackage{lineno}
\usepackage{graphicx}	
\usepackage{multirow,makecell}

\RequirePackage{bm,amsfonts,hhline,amsmath,amssymb,microtype,float,eurosym,bm,
latexsym,epsf,mathtools,cuted,times,makecell,array,natbib,cancel}
\usepackage{tabularx}
\usepackage{epstopdf}
\usepackage[normalem]{ulem}
\usepackage{gensymb}
\usepackage{xcolor}
\usepackage{cancel}

\usepackage{pifont}
\usepackage{ulem}

\usepackage{etoolbox}
\usepackage{environ}
\usepackage{nameref}

\usepackage{orcidlink}

\makeatletter
\NewEnviron{specialalign*}{%
  \appto\@lign{\specialalign@row}%
  \gdef\specialalign@row{\mathversion{normal}}%
  \begin{align*}
  \BODY
  \end{align*}
}
\newcommand{\normalrow}{%
  \noalign{\gdef\specialalign@row{\mathversion{normal}}}%
}
\newcommand{\boldrow}{
  \noalign{\gdef\specialalign@row{\mathversion{bold}}}%
}
\makeatother

\definecolor{ForestGreen}{HTML}{228B22}
\definecolor{deepgreen}{RGB}{0, 100, 0}

\usepackage{graphicx}

\hypersetup{pdfstartview=FitH,pdfhighlight=/O,colorlinks=true}

\begin{document}

\title{Exploring the Impact of Extra Dimensions on Neutron Star Structure and Equation of State}

\author{Debabrata Deb\hspace{0.05cm}\orcidlink{0000-0003-4067-5283}} 
\email{d.deb32@gmail.com}
\affiliation{The Institute of Mathematical Sciences,C.I.T.  campus, Taramani,  Chennai,  600113,  India}

\author{Manjari Bagchi\hspace{0.05cm}\orcidlink{0000-0001-8640-8186}}
\email{manjari@imsc.res.in}
\affiliation{The Institute of Mathematical Sciences,C.I.T.  campus, Taramani,  Chennai,  600113,  India}
\affiliation{Homi Bhabha National Institute, Training School Complex,Anushakti Nagar,  Mumbai, 400094,  India}

\author{Sarmistha Banik\hspace{0.05cm}\orcidlink{0000-0003-0221-3651}}
\email{sarmistha.banik@hyderabad.bits-pilani.ac.in}
\affiliation{Birla Institute of Technology and Science, Pilani Hyderabad Campus, Telangana, India}




\begin{abstract}
In this work, we explore the impact of higher dimensional spacetime on the stellar structure and thermodynamic properties of neutron stars. Utilizing the density-dependent relativistic hadron field theory, we introduce modifications to incorporate the influence of higher dimensionality, a novel approach not explored in existing literature to our best knowledge. Our methodology involves solving the essential stellar structure equations in D-dimensional spacetime ($D \geq 4$), starting with the modification of the Einstein-Hilbert action, derivation of the Einstein field equation in D dimensions, and application of the resulting exterior Schwarzschild spacetime metric for D-dimension. Our findings reveal that with incremental dimensions, the central density $\rho_{c} G_D$ and central pressure $p_c G_D$ gradually increase, leading to progressively stiffer neutron matter. Incremental dimensionality also results in a gradual increase in the maximum mass attained, limited to our study between $D=4$ and $D=6$, as no maximum mass value is obtained for $D>6$. We consistently observe the criteria $dM/d\rho_c>0$ fulfilled up to the maximum mass point, supported by stability analysis against infinitesimal radial pulsations. The validity of our solution is confirmed through causality conditions, ensuring that the matter sound speed remains within the speed of light for all cases. Additionally, our examination indicates that the total mass-to-radius ratio for all discussed D-dimensional cases comfortably resides within the modified Buchdahl limit, which exhibits physical validity of achieved results.

\end{abstract}


\maketitle

\section{Introduction} \label{sect1}

The theory of general relativity (GR) by Einstein stands as one of the remarkable and well-tested theories to broadly understand the phenomenon of gravitational interaction~\cite{Einstein1916,Einstein1918,Hawking2011,Wald1984}. GR has successfully been able to explain results achieved through observations spanning the broad spectrum length scale~\cite{Will2006,Berti2015,Berti2018A,Berti2018B}, such as from the weak gravity field tests like perihelion precession and lensing to the strong gravity filed tests like gravitational waves (GWs) generated through the mergers of compact objects,~\cite{LIGO2017,LIGO2019,LIGO2021} etc. In spite of its great success, GR faced stringent theoretical limitations,  such as the existence of sigularities of spacetime~\cite{Penrose1965,Hawking1976},  the issues related to cosmological constant and late time acceleration~\cite{Weinberg1989,Carroll2001,Padmanabhan2003},  inconsistency between quntum theory and GR~\cite{Birrell1984}, loss of determinism as a result of the voilation of cosmic censorship~\cite{Cardoso2018,Rahman2019}, etc., which urged researchers to come up with an alternative or modified theories.  

The recent decetion of the first multi-messenger GW event, GW170817~\cite{LIGO2017,Abbott2017}, comprising a neutron star (NS) pair and its associated electromagnetic counterpart, has forged a novel pathway for delving into fundamental physics through astrophysical observations. Intriguingly, the GW170817 event has triggered the longstanding notion that GWs may indeed enhance our understanding of astrophysical phenomena within a higher-dimensional framework~\cite{Barvinsky2003,Cardoso2003a,Cardoso2003b,Alesci2005,Cardoso2008}. Numerous researchers have subsequently presented their work~\cite{Yu2017,Andriot2017,Chakraborty2018a,Visinelli2018,Pardo2018,McDonough2018,Liu2023}, demonstrating and discussing the information encoded within GWs pertaining to extra dimensions.  The achieved GW luminosity distance in GW170817 is $40^{+8}_{-14}$ Mpc~\cite{LIGO2017}, whereas the estimated electromagnetic luminosity distance in this event is $40.7^{+2.4}_{-2.4}$ Mpc~\cite{Cantiello2018}, which strongly indicates the possibility that GWs may have traveled through higher dimenional spacetime, whereas electromagnetic waves are restricted to a four-dimensional space~\cite{Andriot2017,Yu2017,Visinelli2018,Lin2020, Du2021,Ishihara2001,Caldwell2001}. Furthermore, this is expected to introduce a time delay between the arrival of GWs and the corresponding light signal observed from the same source associated with the GW170817 event.

Exploring the existence of extra spatial dimensions along with our four-dimensional spacetime is a fascinating endeavor in understanding the fundamental structure of spacetime.  In the pursuit of unifying gravity and electromagnetism in 1914, Gunnar Nordstr{\"o}m initially proposed the concept of an extra spatial dimension~\cite{Nordstrom1914a,Nordstrom1914b}.  A few years later, Theodor Kaluza and Oskar Klein introduced the renowned ``Kaluza-Klein (KK) theory"~\cite{Kaluza1921,Klein1926a,Klein1926b}, which adeptly reinstates both general relativity (GR) and electromagnetism within the four-dimensional spacetime.  In 1983, Valerii A. Rubakov and Mikhail E. Shaposhnikov formulated an alternative higher-dimensional theory known as ``domain wall models"~\cite{Rubakov1983a,Rubakov1983b} by incorporating an infinite extra dimension and a bulk scalar field.   In the realm of string theory, the necessity for ten or more dimensions naturally arises, later resurfaced in addressing the `gauge hierarchy problem'~\cite{Rubakov1983a,Rubakov1983b,Antoniadis1990,Hamed1998,Antoniadis1998,Csaki2004,Perez2005}.  This issue stems from the observed hierarchical and uncorrelated nature of physics scales, particularly the significant disparity between the electro-weak symmetry breaking scale (${10}^3$ GeV) and the Planck scale (${10}^{18}$ GeV). Another practical concern is the fine-tuning, at one part in ${10}^{15}$, required to match the observed Higgs Boson mass at the Large Hadron Collider (LHC), contributing to the continuation of the gauge hierarchy problem~\cite{Csaki2004,Perez2005,Sundrum2005}. In fact subsequent investigations have revealed diverse applications of the higher-dimensional framework in various contexts,  such as black holes~\cite{Dadhich2000,Chamblin2000,Emparan2000,Chamblin2001,Harko2004,Aliev2005,Chakraborty2016,Nakas2021} and cosmology~\cite{Csaki1999,Csaki2000}, etc.

In the realm of gravitational theory, numerous theoretical studies have been persued to comprehend the impact of higher dimensional spacetime on various phenomena associated with compact objects research.  Some investigations~\cite{Harko1993,Harko2000,Leon2000} thoroughly examined the influence of higher-dimensional spacetime on isotropic, anisotropic, and charged fluid spheres. Paul~\cite{Paul2001} established an upper bound for the mass-to-radius ratio of isotropic fluid spheres.  Chakraborty et al.~\cite{Dadhich2017,Chakraborty2018} presented modified Buchdahl limit for various higher-dimensional gravity theories.  Ghosh et al.~\cite{Ghosh2000,Ghosh2001a,Ghosh2001b} explored the formation and characteristics of black holes and naked singularities arising from gravitational collapse within a higher-dimensional framework.  Chavanis~\cite{Chavanis2017} examined the impact of D-dimensional spacetime on relativistic white dwarf stars. This was accomplished through the application of the modified Chandrasekhar equation of state (EoS) derived for D dimensions. Arba{\~n}il et al. studied the hydrostatic equilibrium of strange quark stars through the application of the radial perturbation approach~\cite{Arbanil2019}. Subsequently, employing the Cowling approximation, they explored nonradial perturbation equations for the strange stars influenced by D-dimensional spacetime~\cite{Arbanil2020}.

Motivated by the aforementioned research,  we perform  a comprehensive study of NSs within the context of higher dimensions. The dense matter relevant to NS interior is studied within various phenomenological models. Relativistic Mean Field (RMF) Model with nonlinear scalar meson terms is one of them.  The meson-nucleon couplings are fixed through properties of nuclear matter and finite nuclei at saturation density. The couplings are made density dependent (DD) to take care of the uncertainties at higher densities \cite{Hempel2010, Banik2014}. The density-dependence of the couplings introduces  a rearrangement term in baryon chemical potential and pressure\cite{Banik2021}. In this paper, we have included higher dimension framework to the DD model. Here also the rearrangement term significantly alters the pressure and, consequently, the EoS at higher densities. We explore the impact of additional dimension to the DD model on the structure of NS, a novel aspect, to our best knowledge, not previously investigated in existing literature.

The manuscript is structured as follows: In Section~\ref{sect2}, we present modified Einstein field equations in D-dimensions, essential stellar structure equations, and modified DD model nuclear EoS in D-dimension. The results and graphical representations of different parameters are also discussed and shown in Section~\ref{sect3}. Finally, our work is concluded with a brief discussion in Section~\ref{sect5}. Related detailed mathematical setups are presented in~\ref{appnA} and~\ref{appnB}.

\section{Basic Formalism and Structure Equations for Compact Stars in D Dimension} \label{sect2}

\subsection{D-dimensional Einstein Field Equation}

The generalized Einstein-Hilbert (EH) action in a D-dimensional spacetime parametrized by the coordinates $x^\mu (\mu = 0, 1,  ..., D-1)$ is given by
\begin{equation}\label{1.1}
S=\frac{1}{K_D}\int d^{D}x \mathcal{R} \sqrt{-\mathfrak{g}}+\int d^{D}x\mathcal{L}_m\sqrt{-\mathfrak{g}},
\end{equation}
where $\mathfrak{g}$, $\mathcal{L}_m$, and $\mathcal{R}$ represents determinant of the metric $\mathfrak{g}_{\mu\nu}$, matter lagrangian density, and Ricci scalar, respectively.  Here $K_D$ is the Einstein's constant, which in D-dimensional spacetime takes form as follows
\begin{equation}\label{1.2}
K_D = 2\,\frac{D-2}{D-3}\frac{G_D}{c^4}S_{D-2},
\end{equation}\\
where $G_D$ denotes Newton’s gravitational constant in D-dimension, $c$ is the speed of light.  Here $S_{D-2}$ is the area of the sphere of unit radius in D-dimension, given by
\begin{equation}\label{1.3}
S_{D-2} = \frac{2\,\pi^{\frac{D-1}{2}}}{\Gamma{\left(\frac{D-1}{2}\right)}},
\end{equation}\\
where $\Gamma$ is the gamma function. We will now introduce $T_{\mu\nu}$, which serves as the stress-energy tensor describing the isotropic fluid distribution is given by
\begin{eqnarray}\label{1.4}
T_{\mu\nu}=\left(\rho_D+\frac{p_D}{c^2}\right){u_{\mu}}{u_{\nu}} - {p_D}{\mathfrak{g}_{\mu\nu}},
\end{eqnarray}
where,  $\rho_D$ and $p_D$ denote the density and radial pressure of the matter distribution, while $u_{\mu}$ represents the four-velocity and satisfy $u^\mu u_{\mu}=1$. 

Now variation of EH action of Eq.~\eqref{1.1} and using Eq.~\eqref{1.2} we find the generalized form of the  Einstein field eqution in D-dimension as follows:
\begin{eqnarray}\label{1.5}
\mathcal{R}_{\mu\nu} - \frac{1}{2} \mathcal{R} \mathfrak{g}_{\mu\nu} = \frac{D-2}{D-3} S_{D-2}\frac{G_D}{c^4} T_{\mu\nu},
\end{eqnarray}
where $\mathcal{R}_{\mu\nu}$ is the Ricci tensor.  Evidently, for $D=4$ one may retrieve the well known form of the Einstein field equation, such as $\mathcal{R}_{\mu\nu} - \frac{1}{2} \mathcal{R} \mathfrak{g}_{\mu\nu} =  \frac{8\pi G_4}{c^4} T_{\mu\nu}$.

\subsection{Basic Stellar Equations in D-dimension}
In order to characterize the static hyperspherically symmetric distribution of the fluid, we assume that the interior line element is defined as:
\begin{eqnarray}\label{1.6}
ds^2 = e^{\nu} dt^2 - e^{\lambda} dr^2 -r^2 \sum _{i=1}^{D-2} \left( \prod _{j=1}^{i-1}  \sin^2   \theta_
{{j}}  d\theta^2 _{i} \right),
\end{eqnarray}
where $\nu$ and $\lambda$ serve as metric potentials, both of which are function of the radial coordinate only.

Therefore, by employing Eqs.~\eqref{1.5} and~\eqref{1.6}, we derive the components of Einstein field equation describing the spherically symmetric stellar system within a D-dimensional spacetime, which are presented as follows:
\begin{eqnarray}\label{1.7}
&\qquad\hspace{-2.5cm} {{\rm e}^{-\lambda}} \left[ {\frac { \left( D-2 \right) \lambda^\prime }{2r}}-{\frac { \left( D-2 \right)  \left( D-3 \right) }{2{r}^{2}}} \right] + {\frac { \left( D-2 \right)  \left( D-3 \right) }{2{r}^{2}}}\nonumber\\
&\qquad \hspace{3cm}  =\frac{D-2}{D-3}\frac{S_{D-2} G_D}{c^4} \rho_D c^2,  \\\label{1.8}
&\qquad\hspace{-2.5cm}  {{\rm e}^{-\lambda}} \left[ {\frac { \left( D-2 \right) \nu^\prime}{2r}}+{\frac { \left( D-2 \right)  \left( D-3 \right) }{2{r}^{2}}} \right] - {\frac { \left( D-2 \right)  \left( D-3 \right) }{2{r}^{2}}}\nonumber\\
&\qquad \hspace{3cm}  = \frac{D-2}{D-3}\frac{S_{D-2} G_D}{c^4} p_D,  \\\label{1.9}
&\qquad\hspace{-3cm} {{\rm e}^{-\lambda}} \left[ \frac{\nu^{\prime\prime}}{2}+\frac{{\nu^{\prime}}^{2}}{4}-\frac{\nu^\prime\lambda^\prime}{4}+{\frac { \left( D-3 \right)  \left( \nu^\prime-\lambda^\prime \right) }{2r}} \right] \nonumber\\
&\qquad +{\frac { \left( D-3 \right)  \left( D-4 \right)  \left( {{\rm e}^{-\lambda}}-1 \right) }{{r}^{2}}}=\frac{D-2}{D-3}\frac{S_{D-2} G_D}{c^4} p_D,
\end{eqnarray}
where the symbol prime$\left(\prime\right)$ denotes differentiation with respect to the radial coordinate, $r$.

Now integrating Eq.~\eqref{1.7} for vacuum, i.e., considering $\rho_D = 0$, we have
\begin{eqnarray}\label{2.0}
e^{-\lambda} = 1 -\frac{n}{r^{D-3}},
\end{eqnarray}
where $n$ is the integrating constant.
Conversely, following the principle of correspondence, GR is anticipated to produce results consistent with Newtonian gravity for weak gravitational fields. This requirement entails specifying the ``00" component of the metric tensor, $\mathfrak{g}_{00}$ as
\begin{eqnarray}\label{2.1}
\mathfrak{g}_{00} = 1 + \frac{2\,\Phi}{c^2},
\end{eqnarray}
where $\Phi$ is the gravitational potential.  Again, comparision between Eqs.~\eqref{2.0} and~\eqref{2.1} leads:
\begin{eqnarray}\label{2.2}
\Phi = - \frac{n\,c^2}{2r^{D-3}}.
\end{eqnarray}
As the gravitational field intensity $\mathtt{g} = -\,G_D\,M/r^{D-2}$ is connected to the potential $\Phi$ through the relationship $\mathtt{g} = -\nabla \Phi$, we utilize Eq.~\eqref{2.2} to obtain n, given by
\begin{equation}\label{2.3}
n = \frac{2G_{D}M}{(D-3)c^2},
\end{equation}
where $M$ is the total mass of a star.  Substituting Eq.~\eqref{2.3} into Eq.~\eqref{2.0}, we get
\begin{eqnarray}\label{2.4}
e^{-\lambda} = 1 -\frac{2MG_D}{c^2(D-3)r^{D-3}}.
\end{eqnarray}
Hence, the exterior Schwarzschild spacetime metric in D-dimensional spacetime takes form as follows:
\begin{eqnarray}\label{2.5}
&\qquad\hspace{-1cm}ds^2 = \left(1 -\frac{2MG_D}{c^2(D-3)r^{D-3}}\right)dt^2  -  \left(1 -\frac{2MG_D}{c^2(D-3)r^{D-3}}\right)^{-1}dr^2\nonumber \\
&\qquad\hspace{1cm} -r^2 \sum _{i=1}^{D-2} \left( \prod _{j=1}^{i-1}  \sin^2   \theta_{{j}}  d\theta^2 _{i} \right).
\end{eqnarray}
Based on the form of the exterior Schwarzschild spacetime metric in D-dimensional spacetime let us consider the functional form for $e^{\lambda\left(r\right)}$ as follows
 \begin{eqnarray}\label{2.6}
e^{-\lambda(r)} =  1 -\frac{2m(r)G_D}{c^2\left\lbrace D-3\right\rbrace r^{D-3}}.
\end{eqnarray}
Here $m(r)$, the mass enclosed within the radial coordinate $r$ of the compact star, can be defined using Eqs.~\eqref{1.7} and \eqref{2.5} as
\begin{eqnarray}\label{2.7}
m(r) = \int_0^r  S_{D-2}\, r^{D-2} \rho_D \;\mathrm{d}r.
\end{eqnarray}
Hence, using Eqs. ~\eqref{1.7}-\eqref{1.8} and \eqref{2.7}, the essential stellar structure equations for the compact stars in D-dimensional spacetime reads
\begin{eqnarray}\label{2.8}
&\qquad\hspace{-6cm} \frac{dm}{dr} = S_{D-2} r^{D-2} \rho_D,\\\label{2.9}
&\qquad\hspace{-1.7cm} \frac{d{p_D}}{dr} = -\left(\rho_D c^2 + p_D\right)\frac{G_D\left[S_{D-2} p_D r^{D-1} + c^2  \left\lbrace D-3\right\rbrace m\left(r\right)\right]}{c^2 r \left[c^2 \left\lbrace D-3\right\rbrace r^{D-3} - 2 G_D m\left(r\right)\right]}.
\end{eqnarray}
Eq.~\eqref{2.9} is equivalent to the Tolman-Oppenheimer-Volkoff (TOV) equation. It has been adapted from its original form to account for the impact of generalized D-dimensioanal spacetime~\cite{Leon2000},  which reduces to the conventional TOV equation~\cite{Tolman1939,Oppenheimer1939} for D = 4. For the simplicity of calculation and presentation of the results now onward we will assume that the mass,  density and pressure of a D-dimensional compact star having radius $r$ is denoted as $\tilde{m}\left( r \right)= m\left( r \right) G_D / \left[ D-3 \right]$, $\tilde{\rho} = \rho_D G_D$ and $\tilde{p}=p_D G_D$,  respectively.  Also, we adopt the convention $G_4 = 1$ and $c = 1$.

To determine equilibrium configurations for stars, the stellar structure equations (Eqs.~\eqref{2.8} and \eqref{2.9}) need to be integrated radially from the stellar core to the surface.  To this end, we assume the initial conditions:\\
\begin{eqnarray}\label{3.0}
&\qquad\hspace{-1cm} \tilde{m}(0) =0,  \lambda(0) = 0,  \nu(0) = \nu_c, \nonumber \\ 
&\qquad\hspace{-1cm}   \tilde{\rho}(0) = \tilde{\rho}_c = {G_D \rho_c}~ \text{and}~ \tilde{p}(0) = \tilde{p}_c = {G_D} {p_c},
\end{eqnarray}
where the subscript c stands for center of the star. At the surface the pressure becomes zero, i.e.,  
\begin{equation}\label{3.0.1}
\tilde{p}(R) = 0.
\end{equation}


\subsection{Density Dependent Nuclear EoS in D-Dimension}

The characteristics of atomic nuclei can be effectively elucidated using the mean-field interaction framework. Within this paradigm, the nuclear many-body problem is represented as an energy density functional theory (DFT).  In this study, we employ the DD2 EoS model~\cite{Typel2005, Typel2010} to describe the nuclear matter relevant in NS, consisting of baryonic components like neutrons ($\mathfrak{n}$) and protons ($p$), alongside leptons such as electrons ($e$) and muons ($\bar{\mu}$).  It's important to note that the density dependent Lagrangian incorporates baryon (lepton) fields represented as isospinors denoted as $\psi_B$ ($\psi_l$).  The baryons interact through the exchange of isoscalar-scalar ($\sigma$), Isoscalar-vector ($\omega$) and isovector-vector ($\rho$) mesons. In this work, we have considered that leptons are non-interacting.

Hence, the complete Lagrangian density within the hadronic phase can be expressed as
\begin{eqnarray}\label{3.1}
\mathcal{L} = \mathcal{L}^{fermi} +  \mathcal{L}^{bose} +  \mathcal{L}^{int},
\end{eqnarray}
where 
\begin{align}\label{3.2}
&\qquad\hspace{-1cm} \mathcal{L}^{fermi} = \sum _{B}  \bar{\psi}_B \left[i\gamma^\mu \partial_{\mu} - m_B \right]\psi_B \nonumber \\
&\qquad\hspace{2cm} +  \sum _{l} \bar{\psi}_l \left[i\gamma^\mu \partial_{\mu} - m_l \right]\psi_l , \\ \label{3.3}
&\qquad\hspace{-1cm}   \mathcal{L}^{bose}  = \frac{1}{2}\left(\partial_{\mu}\hat{\sigma} \partial^\mu	 \hat{\sigma} - m^2_\sigma \hat{\sigma}^2 \right) - \frac{1}{4} \hat{\omega}_{\mu\nu} \hat{\omega}^{\mu\nu} + \frac{1}{2} m^2_\omega \hat{\omega}_\mu \hat{\omega}^\mu \nonumber \\
&\qquad\hspace{0cm}  - \frac{1}{4} {\bm{\hat{\rho}_{\mu\nu}}}\cdot {\bm{\hat{\rho}^{\mu\nu}}} + \frac{1}{2} m^2_\rho  {\bm{\hat{\rho}_{\mu}}}\cdot {\bm{\hat{\rho}^{\mu}}}  \nonumber \\
&\qquad\hspace{0cm}, \\ \label{3.4}
&\qquad\hspace{-1cm} \mathcal{L}^{int} = \sum _{B} \bar{\psi}_B \big[ \hat{g}_{\sigma B}\hat{\sigma} - \hat{g}_{\omega B} \hat{\omega}_\mu \gamma^\mu \nonumber \\
&\qquad\hspace{2cm} - \hat{g}_{\phi B} \hat{\phi}_\mu \gamma^\mu   - \hat{g}_{\rho B} \bm{\tau_B \cdot \hat{\rho}_\mu } \gamma^\mu  \big],
\end{align}
where $\bm{\tau_B}$ is the isospin operatore and we define the field strength tensors for vector mesons as following
\begin{eqnarray}\label{3.5}
&\qquad \hat{\omega}_{\mu\nu} = \partial_\mu \hat{\omega}_\nu - \partial_\nu \hat{\omega}_\mu,\\\label{3.6}
&\qquad \hat{\phi}_{\mu\nu} = \partial_\mu \hat{\phi}_\nu - \partial_\nu \hat{\phi}_\mu,\\\label{3.7}
&\qquad \hat{\bm{\rho}}_{\mu\nu} = \partial_\mu \hat{\bm{\rho}}_\nu - \partial_\nu \hat{\bm{\rho}}_\mu.
\end{eqnarray}
Here the symbols $\hat{g}_{\alpha B}(\hat{n})$ represent the interaction strengths between mesons and baryons and are dependent on vector density. Here, $\alpha$ can take on values such as $\sigma$, $\omega$,  and $\rho$. The density operator $\hat{n}$ is defined as $\hat{n} = \sqrt{\hat{j}_\mu\hat{j}^\mu}$, where $\hat{j}^\mu = \bar{\psi} \gamma^\mu \psi$. 

As $\hat{g}_{\alpha B}$ are Lorentz scalar functionals of baryon field operators, the derivative of $\mathcal{L}$ with respect to $\bar{\psi}_B$ yields the following:
\begin{equation}\label{3.8}
\frac{\delta \mathcal{L}}{\delta \bar{\psi}_B} = \frac{\partial \mathcal{L}}{\partial \bar{\psi}_B} + \frac{\partial \mathcal{L}}{\partial \hat{\rho}_B} \frac{\partial \hat{\rho}_B}{\partial \bar{\psi}_B}
\end{equation}
Additionally, the meson-baryon coupling strength becomes a function of the total baryon density $n$, denoted as $\left\langle \hat{g}_{\alpha B} (\hat{n})\right\rangle = g_{\alpha B} (\left\langle \hat{n} \right\rangle) = g_{\alpha B} (n)$.

The meson field equations are solved in a self-consistent manner, taking into account the charge neutrality and the conservation of baryon number.  In static and isotropic system all the space and time derivatives of the fields vanish. Furthermore, in the matter's rest frame, the spatial components of $\omega_\mu$ and $\rho_\mu$ become zero.  In the ground state, the interaction between baryons and the third component of the isovector $\rho$ meson arises as the expectation values of sources become zero. RMF approximation is employed to solve the meson field equations, where the expectation values replace the respective meson fields. These equations for the meson fields are expressed as follows:
\begin{align}\label{3.9}
&\qquad \hspace{-1cm} m_\sigma^2  \sigma = \sum _{B} g_{\sigma B}  \rho^{s}_{DB}, \\ \label{4.0}
&\qquad \hspace{-1cm} m^2_\omega \omega_0 =  \sum _{B} g_{\omega B} \rho_{DB}, \\ \label{4.1}
&\qquad \hspace{-1cm}  m^2_\rho \rho_{03} =  \sum _{B} g^2_{\rho B} \tau_{3B} \rho_{DB},
\end{align}
where the variables $ \rho^{s}_{DB}$ and $\rho_{DB}$ blue represent the scalar number density and number density for each baryon,  respectively. They are defined as:
\begin{eqnarray}\label{4.2}
&\qquad \rho^{s}_{DB} = \frac{2J_B+1}{(2\pi)^{D-1}} S_{D-2} \int_0^{k_B} \frac{k^{D-2} m_B^\star}{\sqrt{k^2+{m_B^\star}^2}} dk,\\ \label{4.3}
&\qquad\hspace{-0.8cm}  \rho_{DB} = \frac{2J_B+1}{(2\pi)^{D-1}} S_{D-2} \int_0^{k_B} k^{D-2} dk.
\end{eqnarray}
Here,  $J_B$ typically represents the total angular momentum quantum number associated with the baryon $B$. Here $m^\star_B$ is the effective baryon mass,  given by
\begin{eqnarray}\label{4.4}
m^\star_B = m_B - g_{\sigma B}\sigma.
\end{eqnarray}
In the framework of RMF approximations, the Dirac equations for the baryon fields can be expressed as follows:
\begin{eqnarray}\label{4.5}
\left[\left(i\gamma^\mu \partial^\mu - \gamma^0 \Sigma^\tau_0 \right) - m^\star_B \right] \psi_B = 0,
\end{eqnarray}
Here, the symbol $\Sigma^\tau_0$,  the rearrangement term is
\begin{eqnarray}\label{4.6}
\Sigma^\tau_0 = g_{\omega B} \omega_0 +  g_{\phi B} \phi_0 + g_{\rho B} \tau_{3B} \rho_{03} + \Sigma^R_0,
\end{eqnarray}


\begin{figure*}[t]
\centering
    \includegraphics[width=8cm]{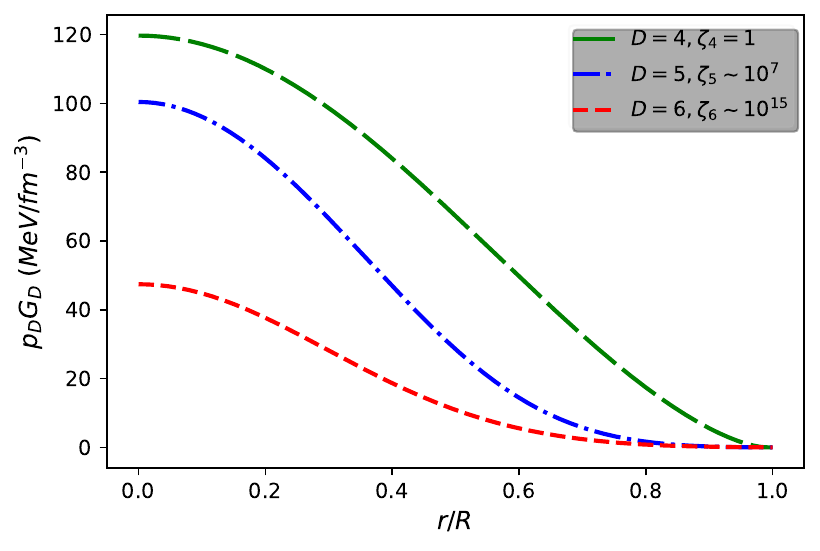}
	\includegraphics[width=8cm]{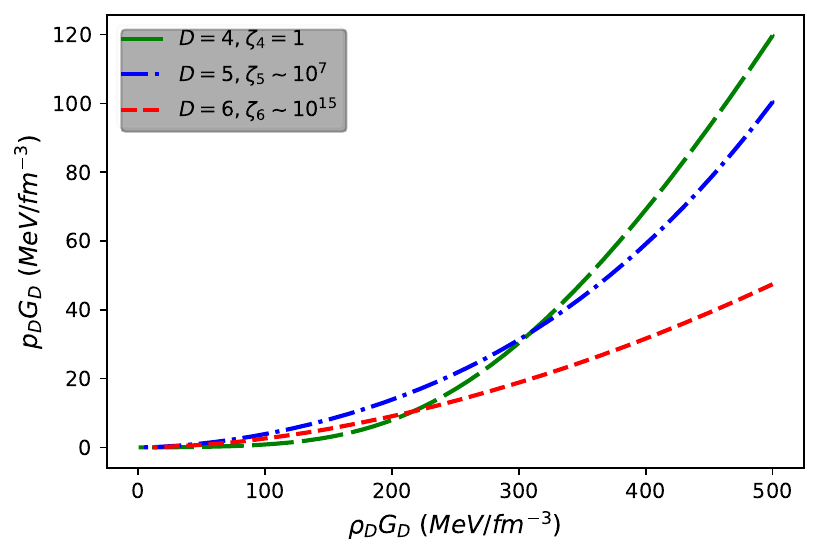}
		\caption{Variation of i) pressure $(\tilde{p} = {p_D G_D})$ with the normalized radial coordinate $(r/R)$ (upper panel) and  ii) pressure $(\tilde{p} = {p_D G_D})$ with the  density~$(\tilde{\rho} = {\rho_D G_D})$ (lower panel) for the different D-dimensional cases, such as $D=4, 5~ \text{and}~6$} \label{fig_eos}
\end{figure*}


where $\Sigma^R_0$ takes the form as given by:
\begin{eqnarray}\label{4.7}
\hspace*{-1cm} \Sigma^R_0 = \sum_B \left[-{g_{\sigma B}}^\prime  \sigma \rho^{s}_{BD} + {g_{\omega B}}^\prime \omega_0  \rho_{DB} + {g_{\rho B}}^\prime \tau_{3B} \rho_{03} \rho_{DB}   \right]. \nonumber
\hspace*{-1cm} \\
\end{eqnarray}
Note that in Eq.~\eqref{4.7}, $g_{\alpha B}'=\frac {\partial g_{\alpha B}} {\partial \rho_B}$, where $\alpha= \sigma,~ \omega, ~\rho$, and $\tau_{3B}$ represents the isospin projection of $B$, which for the present work is $n$ and $p$.

Therefore, the energy density denoted as $(\rho_D)$, and the pressure $(p_D)$ of NS matter under the conditions of $\beta$-equilibrium and local charge neutrality condition, considering a quantum hydrodynamic description for the nuclear matter comprising nucleons, mesons, along with leptons, can be expressed as follows:
\begin{widetext}
\begin{align}\label{4.8}
&\qquad \hspace{-1cm} \rho_D = \frac{1}{2} m^2_\sigma {\sigma}^2 +  \frac{1}{2} m^2_\omega {\omega_0}^2 + \frac{1}{2} m^2_\rho {\rho_{03}}^2  + S_{D-2} \sum_{B} \frac{2J_B+1}{(2\pi)^{D-1}}  \int_0^{k_B} k^{D-2} \sqrt{k^2+{m_B^\star}^2} dk \nonumber\\
&\qquad\hspace*{6cm} + S_{D-2}  \sum_{l} \frac{2J_l+1}{(2\pi)^{D-1}} \int_0^{k_l} k^{D-2} \sqrt{k^2+m^2_l} dk, \\
\text{and} \nonumber\\\label{4.9}
&\qquad \hspace{-1cm} p_D =- \frac{1}{2} m^2_\sigma {\sigma}^2 +  \frac{1}{2} m^2_\omega {\omega_0}^2 + \frac{1}{2} m^2_\rho {\rho_{03}}^2 + {\Sigma^R_0} \sum_{B} \rho_{DB}  + \frac{S_{D-2}}{D-1}  \sum_{B} \frac{2J_B+1}{(2\pi)^{D-1}}  \int_0^{k_B} \frac{2(D-2)}{\sqrt{k^2+{m_B^\star}^2}} dk \nonumber\\
&\qquad\hspace*{6cm} + \frac{S_{D-2}}{D-1}  \sum_{l} \frac{2J_l+1}{(2\pi)^{D-1}}  \int_0^{k_l} \frac{2(D-2)}{\sqrt{k^2+m_l^2}} dk.
\end{align}
\end{widetext}

The nucleon-meson density-dependent couplings $\left(g_{\alpha B}(n)\right)$ are calculated using the methodology outlined by Typel et al. \cite{Typel2005, Typel2010}. The formulation of these couplings with respect to density was initially introduced by Teypel and Wolter~\cite{Typel1999} and can be characterized as follows:
\begin{equation}\label{5.0}
g_{\alpha B}(n) = g_{\alpha B}(n_{0}) f_{\alpha}(x).
\end{equation}
Here, $n$ represents the total baryon density, defined as $n = \sum_B \rho_{DB}$, and $x = n/n_0$. The functional forms of $f_{\alpha}(x)$ vary depending on the meson type.  For the further details one may look into the existing related works~\cite{Typel1999,Typel2005,Typel2010,Hempel2010, Banik2014,Banik2021} in literature and references within.

The masses of the neutron and proton are approximately $939.56536$ MeV and $938.27203$ MeV, respectively, while the $\omega$ and $\rho$ mesons have masses of $783$ MeV and $763$ MeV, as documented in Table II of their work by Typel et al.~\cite{Typel2010}.\\


\begin{figure}[!htpb]
\centering
	\includegraphics[width=8cm]{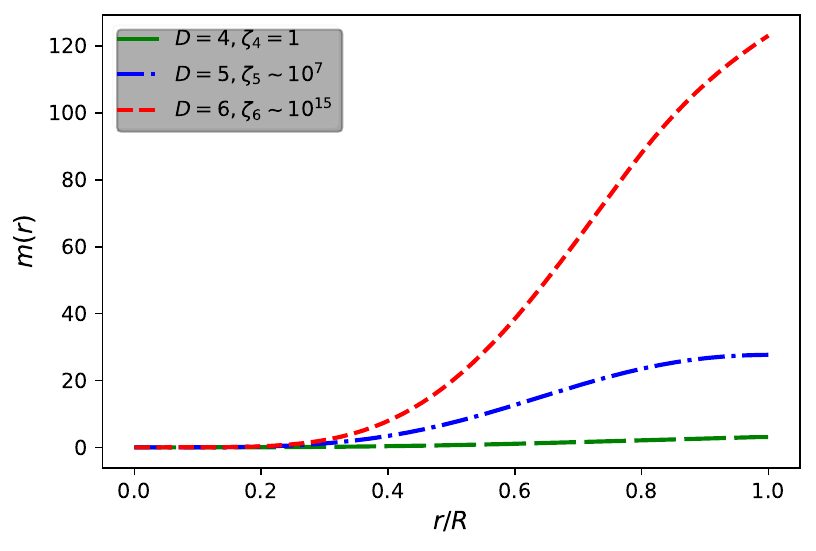}
		\caption{Variation of normalized enclosed mass function $m(r)/M$ with the normalized radial coordinate $(r/R)$ for the different D-dimensional cases, such as $D=4, 5~ \text{and}~6$ considering $\rho_c G_D = 500~ \text{MeV/fm}^3$. Here, the total enclosed mass, $\tilde{m}$ (in geometrized unit) is $3.08$ km for $D=4$,  $27.68~\text{km}^2$ for $D=5$ and $123.14~\text{km}^3$ for $D=6$. } \label{fig_mass}
\end{figure}



\begin{figure}[!htpb]
\centering
	\includegraphics[width=8cm]{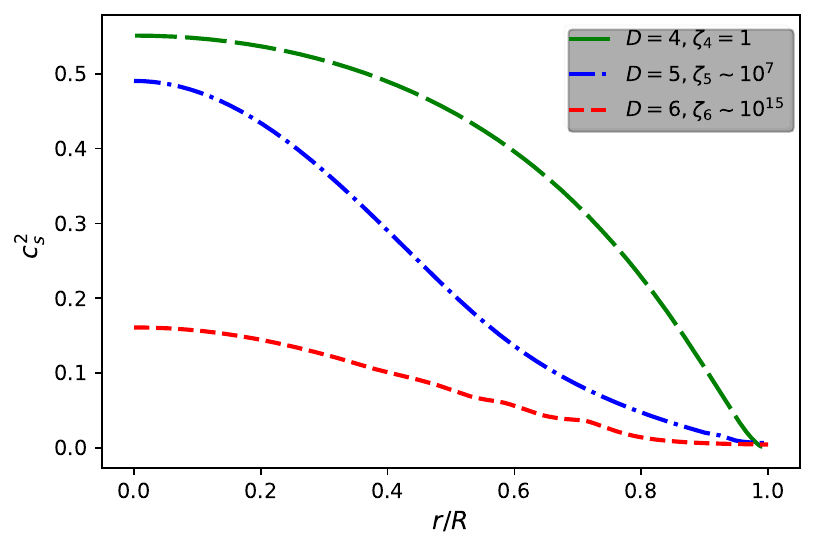}
		\caption{Variation of square of sound speed ($c^2_s$) with radial coordinate $r/R$ along for the different D-dimensional cases, such as $D=4, 5~ \text{and}~6$, each with $\rho_c G_D = 500~ \text{MeV/fm}^3$.} \label{fig_svel}
\end{figure}


\begin{figure*}[t] 
\centering
	\includegraphics[width=5.5cm,height = 4cm]{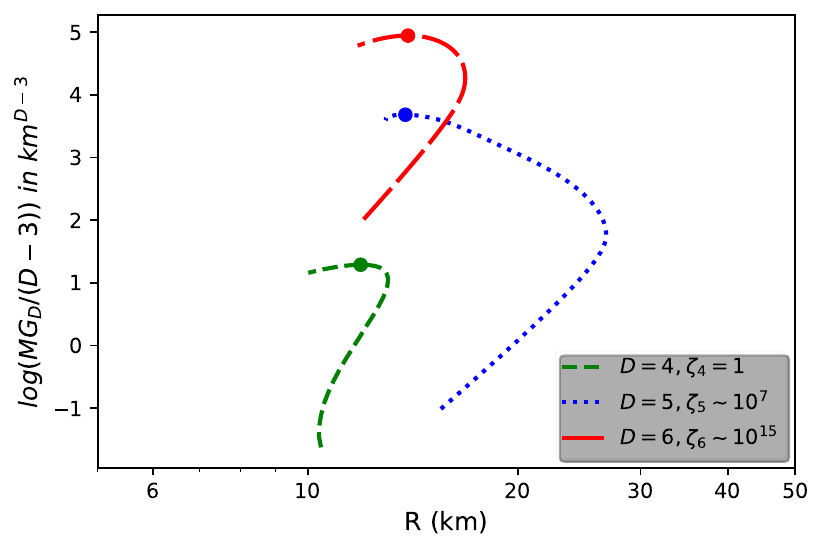}
	\includegraphics[width=5.5cm,height = 4cm]{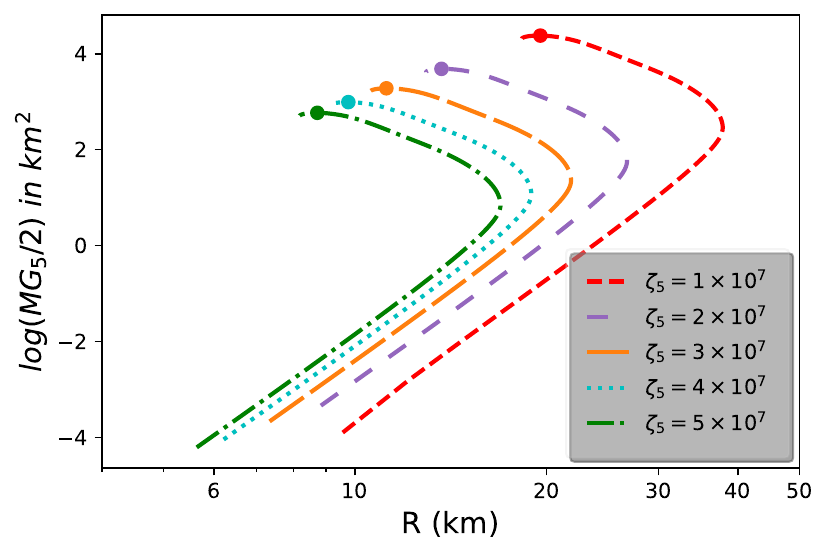}
	\includegraphics[width=5.5cm,height = 4cm]{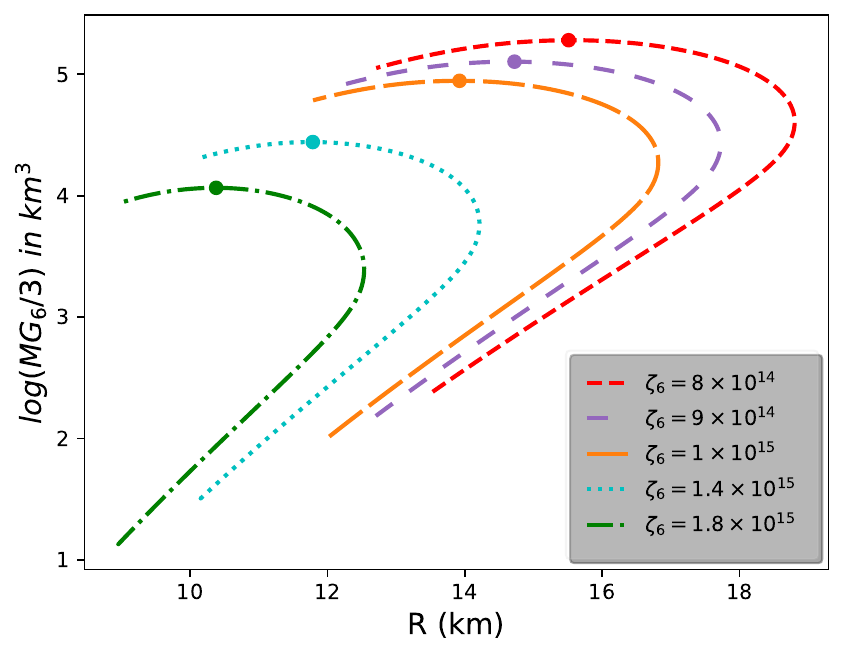}
		\caption{Variation of i) the total normalized mass $({\rm{log}}(M{G_D}/(D - 3)))$ in km$^{(D-3)}$ with the total radius $R$ in km for the different $D$-dimensional cases, such as $D=4, 5~ \text{and}~6$ (left panel), ii) the total normalized mass $({\rm{log}}(M{G_5}/2))$ in km$^{2}$ with the total radius $R$ in km for $D = 5$ for different values of $\zeta_5$ (middle panel) and iii) the total normalized mass $({\rm{log}}(M{G_6}/3))$ in km$^{3}$ with the total radius $R$ in km for $D =  6$ for different values of $\zeta_6$ (right panel). The solid circles in each of the lines denote the maximum mass point.} \label{fig_MR}
\end{figure*}

\begin{figure*}[!htpb]
\centering
	\includegraphics[width=5.5cm,height = 4cm]{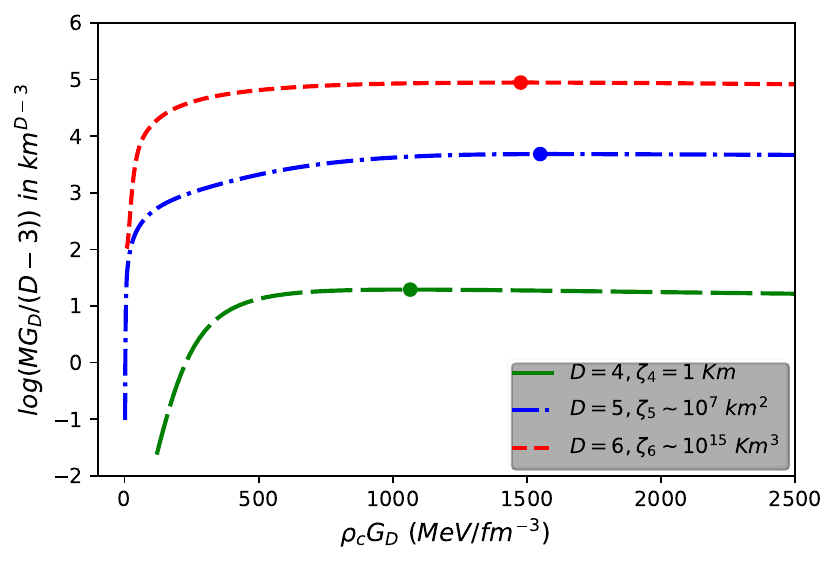}
	\includegraphics[width=5.5cm,height = 4cm]{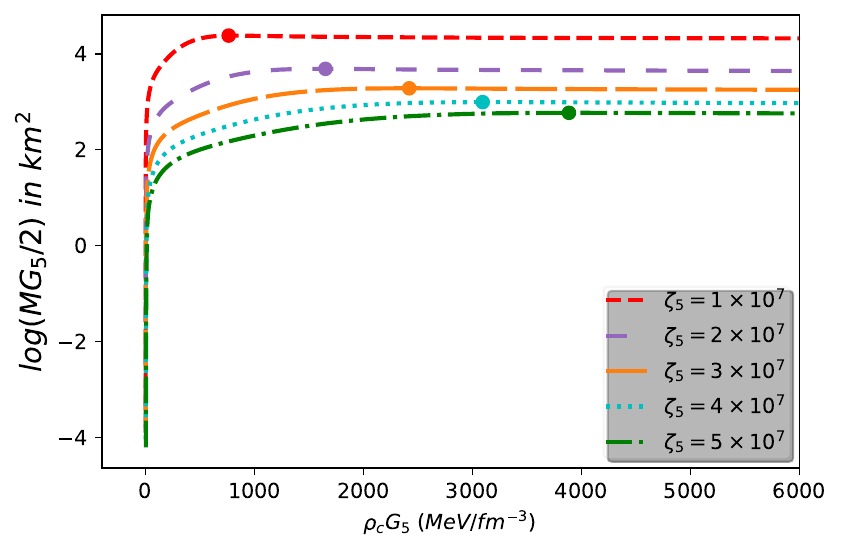}
	\includegraphics[width=5.5cm,height = 4cm]{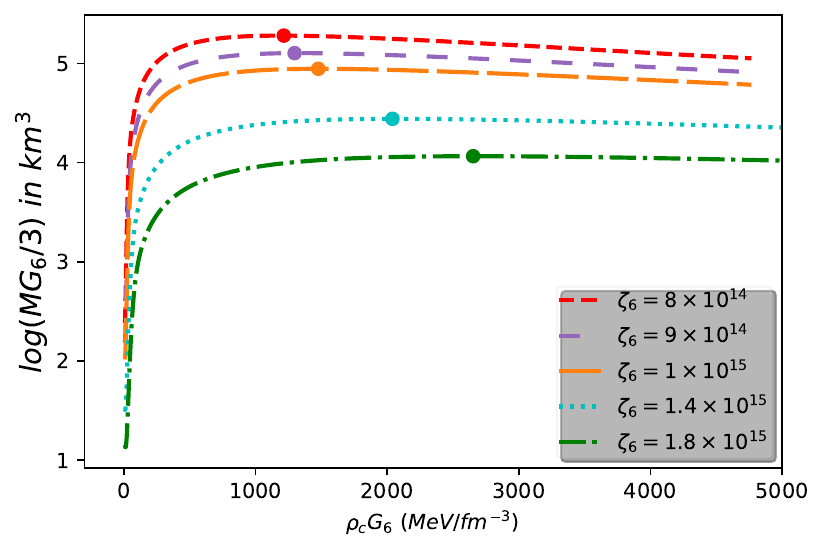}
		\caption{Variation of i) the total normalized mass $({\rm{log}}(M{G_D}/(D - 3)))$ in km$^{(D-3)}$ with the central density $\tilde{\rho}_c=\rho_c G_D$ in $\text{MeV/fm}^3$  for the different $D$-dimensional cases, such as $D=4, 5~ \text{and}~6$ (left panel), ii) the total normalized mass $({\rm{log}}(M{G_5}/2))$ in km$^{2}$ with the centra density $\rho_c G_5$ in $\text{MeV/fm}^3$ for $D = 5$ for different values of $\zeta_5$ (middle panel) and iii) the total normalized mass $({\rm{log}}(M{G_6}/3))$ in km$^{3}$ with the centra density $\rho_c G_6$ in $\text{MeV/fm}^3$ for $D = 6$ for different values of $\zeta_6$ (right panel).  The solid circles in each of the lines denote the maximum mass point.} \label{fig_Mcden}
\end{figure*}

\section{Results and Discussions}\label{sect3}

In this article, our focus is on investigating the influence of higher dimensions on the structural characteristics and nuclear EoS of NSs. To accomplish this, we have applied the modified Einstein field equation (Eq. ~\eqref{1.5}) applicable for D-dimensional spacetime. Subsequently, we have presened the fundamental equations governing the structural properties of NSs (Eq. \eqref{2.8}). In our pursuit to address these structural equations, wWe utilized the modified density-dependent relativistic hadron field theory with the DD2 model EoS to describe matter distribution in NSs in D dimensions.  This adaptation has led us to formulate a generalized functional representation for the energy density, denoted as $\rho_D$, and the pressure, denoted as $p_D$, in the context of D-dimensional spacetime (as detailed in Eqs.~\eqref{4.8} and \eqref{4.9}).

In order to derive EoS, we employ a self-consistent solution of the equations of motion governing mesons and baryons (Eqs.~\eqref{3.9}-\eqref{4.1} and~\eqref{4.5}). These equations are solved within the mean-field approximation, considering additional constraints such as the conservation of baryon number, charge neutrality, and beta equilibrium.  Note that the gravitational constants $G_4$, $G_5$, $G_6$, and so on, may not necessarily be equal to one another, and furthermore, we don't know the values for these constants in various dimensions (D), except $G_4$. To address this, we employ dimensional analysis to introduce a scaling parameter, $\zeta_D$, such that $G_D/G_4 = \zeta_D \,\text{km}^{D-4}$ in geometrized unit. This parameter, $\zeta_D$, remains a free constant. Upon our analysis, we have determined that setting $\zeta_5 \sim 10^7$ and $\zeta_6 \sim 10^{15}$ produces solutions to the stellar structure equations. Deviating by one order of magnitude, either higher or lower, in the chosen parametric values for $\zeta_5$ and $\zeta_6$ renders these solutions unattainable.

To establish hydrostatic equilibrium and tackle the stellar structure equations (Eqs. \eqref{2.8} and \eqref{2.9}), we utilize a numerical approach that integrates EoS described by Eqs.~\eqref{4.8} and \eqref{4.9}, in conjunction with the boundary conditions specified by Eqs.~\eqref{3.0} and~\eqref{3.0.1}. This computational method leverages the fourth-order Runge-Kutta technique along with the shooting method. Our methodology involves a stepwise numerical solution of Eqs.~\eqref{4.8} and \eqref{4.9} as we traverse from the center to the surface of the star. Throughout the neumerical analyisis, we systematically vary parameters such as $D$ and $\zeta_D$ to illustrate the impact of higher dimensionality, and $\rho_c G_D$ to attain solutions for Eqs. \eqref{2.8} and \eqref{2.9}. Importantly, we rigorously uphold the boundary condition Eq.~\eqref{3.0.1} throughout this computation.


\begin{figure}[!htpb]
\centering
	\includegraphics[width=8cm]{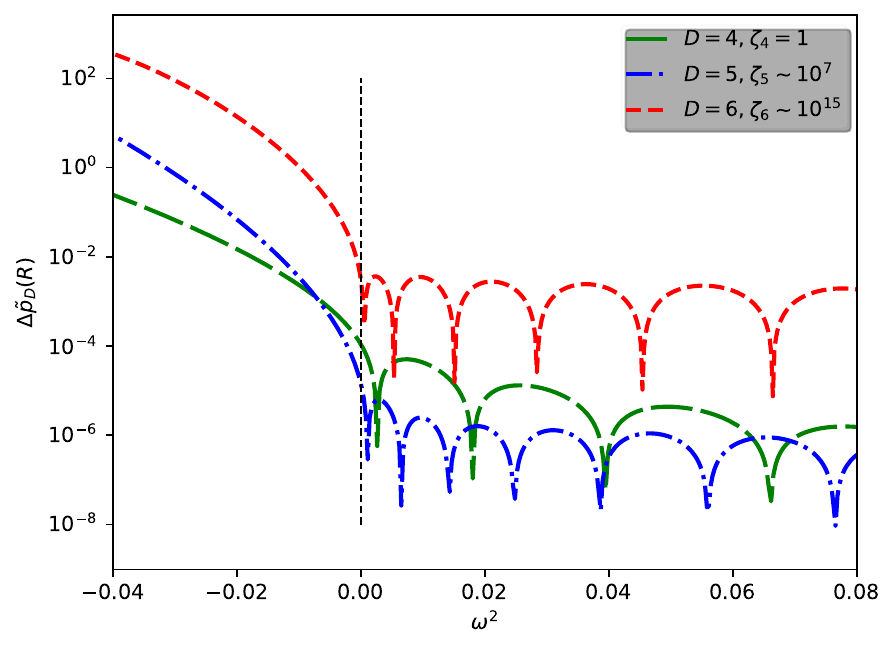}
		\caption{The logarithmic representation of the Lagrangian perturbation in radial pressure ($\Delta \tilde{p}_D(R) = \Delta p_D(R) G_D$) at the outermost layer of a star is depicted along the vertical axis for various trial values of $\omega^2$ along the horizontal axis, while maintaining a constant central mass density $\rho_c G_D = 500~ \text{MeV/fm}^3$  for the different D-dimensional cases, such as $D=4, 5~ \text{and}~6$. Each curve exhibits minima, pinpointing the precise frequencies of oscillation modes within equilibrium configurations. It is noteworthy that since the eigenfrequency of the fundamental mode agrees $\omega^2_0 > 0$, all the NSs for the different D-dimensional cases, such as $D=4, 5~ \text{and}~6$ exhibit stability against radial oscillations.} \label{fig_stability}
\end{figure}



\begin{figure}[!htpb]
\centering
	\includegraphics[width=8cm]{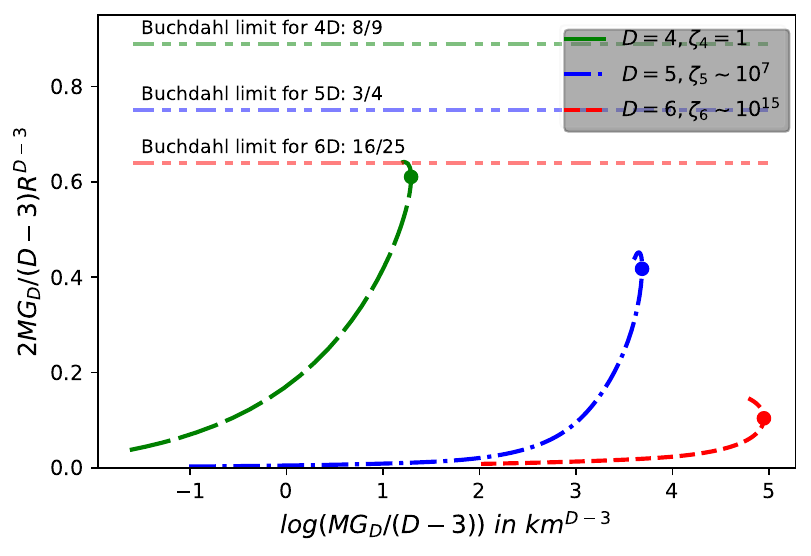}
	\includegraphics[width=8cm]{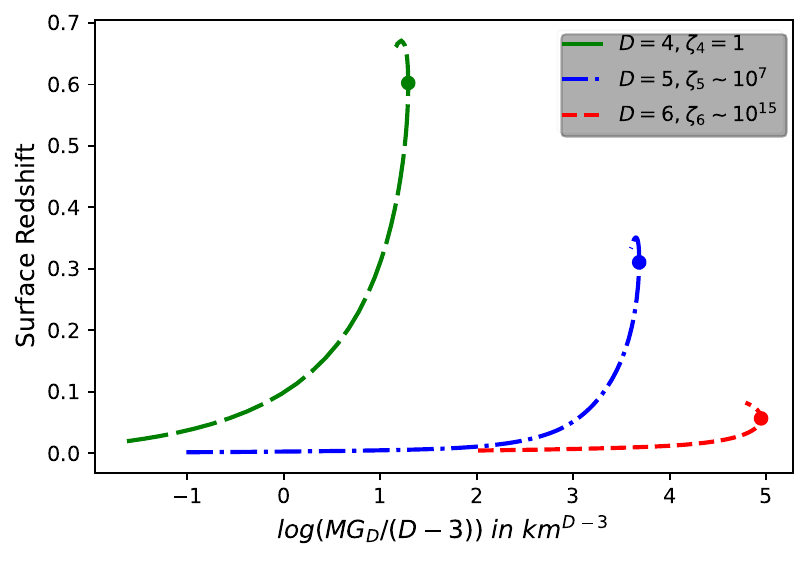}
		\caption{Variation of i) compactnes $2M G_D/(D-3) R^{D-3}$ along the vertical axis with the normalized total mass $({\rm{log}}(M{G_D}/(D - 3)))$ along the horizontal axis (upper panel) and ii) surface redshift along the vertical axis with the normalized total mass $({\rm{log}}(M{G_D}/(D - 3)))$ along the horizontal axis (lower panel) for the different D-dimensional cases, such as $D=4, 5~ \text{and}~6$} \label{fig_redshift}
\end{figure}


In order to present and compare our findings consistently across various dimensions ($D$), we have chosen a reference central density of $\tilde{\rho}_c = 500~\text{MeV/fm}^3$. This choice ensures comparability across dimensions $4$, $5$, and $6$, as they commonly share this reference density. This choice allows us to investigate how various internal properties of NSs change as we transition to higher-dimensional spacetimes. In the upper panel of Fig.~\ref{fig_eos}, we illustrate the pressure profile, which exhibits a gradual decrease from its maximum value at the center to reach a minimum value, $\tilde{p}(R) = 0$, at the surface. Notably, the central pressure decreases as we move from lower to higher dimensions.  Similarly, in the lower panel of Fig.~\ref{fig_eos},  we visualize EoS profiles. The profiles establish the fact that the stiffness of the EoS decreases as the dimensionality of the spacetime increases. This comparative analysis allows us to discern how the central pressure and EoS characteristics evolve with respect to dimensionality.  In Fig.~\ref{fig_mass}, we illustrate the variation of the enclosed mass, $\tilde{m}(r)$ along the vertical axis, with the normalized radial coordinate $r/R$ along the horizontal axis across different dimensions. Our findings reveal a gradual increase in the total mass for a constant $\tilde{\rho_c}$ as we transition to higher dimensions.  For the chosen central density of $\tilde{\rho_c} = 500~\text{MeV/fm}^3$, we find the enclosed mass values as $3.08~\text{km}$, $27.68~\text{km}^2$, and $123.14~\text{km}^3$ corresponding to the respective dimensions $D=4$, $D=5$, and $D=6$. To ensure the physical validity of the modified DD2 model EoS, we assessed its consistency with the causality condition, primarily by examining the square of the sound speed, $c_s^2$, as a function of the radial coordinate $r/R$, as depicted in Fig.~\ref{fig_svel}. Our results demonstrate that for all three cases (i.e., $D=4$, $D=5$, and $D=6$), the square of the sound speed, $c_s^2$, remains below the speed of light, reaffirming the compliance of EoS with the causality condition.  

We study the relationship between their normalized total mass $(\tilde{M}={\rm{log}}\left\lbrace M{G_D}/(D - 3)\right\rbrace)$ and total radius $R$. In the left panel of Fig.~\ref{fig_MR},  we illustrate $\tilde{M}$ versus total radius relationship for NSs across various $D$-dimensional spacetimes, where the solid circles indicate the maximum mass points.  Fig.~\ref{fig_MR} is important for understanding results in higher dimensions. It helps compare with DD2 models in $D=4$. We use it to study how higher dimensions affect compact stars and their properties. With the parameter choices $\zeta_4=1$, $\zeta_5 \sim 10^7$ and $\zeta_6 \sim 10^{15}$, we find that the maximum masses achieved are $3.63~\text{km}$, $39.80~\text{km}^2$, and $140.66~\text{km}^3$ for dimensions $D=4$, $D=5$, and $D=6$, respectively,  accompanied by corresponding radii of $11.908~\text{km}$, $13.803~\text{km}$, and $13.923~\text{km}$, respectively.  The corresponding value of central density, $\tilde{\rho}_c$ for maximum masses are $1064.85~\text{MeV/fm}^{3}$,  $1549.21~\text{MeV/fm}^{3}$,  and $1476.45~\text{MeV/fm}^{3}$, respectively. In the middle and right panels of Fig.~\ref{fig_MR}, we display a series of mass-radius curves for various choosen parametric values of $\zeta_D$, corresponding to $D = 5$ and $D = 6$. In both higher-dimensional scenarios, it becomes evident that as $\zeta_D$ increases, there is a progressive decrease in the maximum mass value and the associated radius for NSs. Our study is limited to $D$ values between 4 and 6. For $D>6$, we do not observe a maximum mass point. 

We present the relationship between normalized total mass $(\tilde{M})$, and central density $\left(\tilde{\rho}\right)$ for different $D$ dimensional cases, in the left panel of Fig.~\ref{fig_Mcden}, which clearly demonstrates that the criterion $d{M}/d{\rho_c} > 0$ is upheld up to the maximum mass point. Interestingly, Fig.~\ref{fig_Mcden} also highlights that the stability criterion $d{M}/d{\rho_c} > 0$, which is the known condition for stable stellar structure in D = 4.  In the middle and right panels of Figure~\ref{fig_Mcden}, our analysis illustrates that the stability criterion holds true for all selected parametric values of $\zeta_D$ in both $D = 5$ and $D = 6$ dimensions, extending up to the maximum mass point. Additionally, as $\zeta_D$ increases, we observe that the maximum mass points are reached at higher central densities. This observation suggests that in a higher-dimensional spacetime, an increase in $\zeta_D$ corresponds to NSs adopting a denser and more compact structural configuration.

After determining the equilibrium quantities through the solution of the TOV equations, the next step involves assessing the stability of higher-dimensional stellar systems against infinitesimal radial pulsations (see \ref{appnA} for details). This entails solving the system of coupled first-order equations~\eqref{A1.2} and \eqref{A1.3} while adhering to the specified boundary conditions~\eqref{A1.4} and \eqref{A1.5}. We employ the shooting method for the numerical solution of these equations, integrating them over a range of trial values for $\omega^2$ that meet the condition~\eqref{A1.4}, where $\omega$ is the eigenfrequency. Additionally, we ensure that the normalized eigenfunctions correspond to $\xi(0) = 1$ at the stellar center and integrate them out to the stellar surface, where $\xi$ denotes the relative radial displacement. The squared frequencies that satisfy the boundary condition ~\eqref{A1.5} represent the correct frequencies of radial oscillations.  Specifically, for $\tilde{\rho}_c=500~\text{MeV/fm}^3$, Figure~\ref{fig_stability} illustrates the Lagrangian perturbation of the radial pressure for a range of test values $\omega^2$ within different $D$-dimensional spacetimes. Each minimum point on Figure~\ref{fig_stability} corresponds to the respective appropriate frequency. Notably, Figure~\ref{fig_stability} demonstrates that our current research results in stable stellar configurations against infinitesimal radial pulsations across the selected range of higher-dimensional spacetimes.

The Buchdahl limit~\cite{Buchdahl1959} imposes a restriction on the maximum compactness of NSs and preventing their gravitational collapse into black holes. We present in detail the modified Buchdahl limit for D dimension in~\ref{appnB}. We illustrate the variation of compactness $\tilde{M}/R =2M G_D/(D-3) R^{D-3} $ with $\tilde{M}$ in the upper panel of Fig.~\ref{fig_redshift} for $D =4,  5~\text{and}~6$. It is evident that for all D-dimensional cases, the masses comfortably remain within the upper limit predicted by the modified Buchdahl limit as given in Eqn.~\eqref{B2.6}. The surface redshift in D dimension takes form as $e^{-\nu{\left(R\right)}/2} - 1 = 1/\sqrt{1-2M G_D/(D-3) R^{D-3} } -1$. In the lower panel of Fig.~\ref{fig_redshift}, we show the variation of surface redshift with $\tilde{M}/R$ for different D-dimensional scenarios. Our findings reveal that with increasing dimensions, the surface redshift of NSs gradually decreases.

In Table~\ref{Table 1}, we present values of physical parameters for the maximum mass of NS in different D-dimensional cases. Tables~\ref{Table 2} and \ref{Table 3} show the changes in physical parameters for $D=5$ and $D=6$, respectively, with varying $\zeta_5$ and $\zeta_6$. The maximum mass is achieved for higher $\tilde{\rho}_c$ values in $D>4$ cases. Increasing $\zeta_5$ and $\zeta_6$ leads to a gradual stiffening of the NS EoS for $D=5$ and $D=6$ cases, respectively. The results are consistent with the modified Buchdahl limit $\frac{M G_D}{c^2 \left(D - 3\right) R^{D - 3}} < \frac{2 \left(D - 2\right)}{\left(D - 1 \right)^2}$ (see \ref{appnB} for details) for all cases.


\begin{table*}[htbp!]
  \centering
    \caption{Numerical values of physical parameters for the NSs  for different values of $\rm D$}
\label{Table 1}
        \begin{tabular}{cccccccccccccccccccc}
\hline\hline  Value  & Value of Maximun & Value of  & Central & Central & & Surface & \\  
of $D$  &  Mass $M{G_D}/(D - 3)$ & Corresponding  & Density & Pressure & $\frac{2M G_D}{ \left(D - 3\right) R^{D - 3}}$ & Redshift & \\ 
  &  in km$^{(D-3)}$  & Radius (km)  & $\widetilde{\rho}_c~(\rm MeV/{fm}^3)$ & $\widetilde{p}_c~(\rm MeV/{fm}^3)$ &  &   & 
\\ \hline 
 $4$ & $3.63$ & $11.91$ & $1064.85$ & $501.78$ & $0.61$ & $0.60$   \\ 
$5$ & $39.80$ & $13.80$ & $1549.20$ & $2671.00$ & $0.42$ & $0.31$\\
$6$ & $140.66$ & $13.92$ & $1476.45$ & $342.56$ & $0.10$ & $0.06$\\
  \hline\hline
  \end{tabular}
    \end{table*}



\begin{table*}[htbp!]
  \centering
    \caption{Numerical values of physical parameters for the NSs  for different values of  $\zeta_5$ and $\rm D=5$}
\label{Table 2}
        \begin{tabular}{cccccccccccccccccccc}
\hline\hline  Value  & Value of Maximun & Value of  & Central & Central & & Surface & \\  
of $\zeta_5$  &  Mass $M{G_D}/(D - 3)$ & Corresponding  & Density & Pressure & $\frac{2M G_D}{ \left(D - 3\right) R^{D - 3}}$ & Redshift & \\ 
  &  in km$^{(D-3)}$  & Radius (km)  & $\widetilde{\rho}_c~(\rm MeV/{fm}^3)$ & $\widetilde{p}_c~(\rm MeV/{fm}^3)$ &  &   & 
\\ \hline 
 $1\times{10}^7$ & $79.58$ & $19.57$ & $762.62$ & $1275.67$ & $0.42$ & $0.31$   \\ 
$2\times{10}^7$ & $39.79$ & $13.68$ & $1650.83$ & $3187.36$ & $0.43$ & $0.32$\\
$3\times{10}^7$ & $26.53$ & $11.20$ & $2419.71$ & $4491.91$ & $0.42$ & $0.32$\\
$4\times{10}^7$ & $19.90$ & $9.76$ & $3093.36$ & $5316.66$ & $0.42$ & $0.31$\\
$5\times{10}^7$ & $15.92$ & $8.73$ & $3885.26$ & $6738.97$ & $0.42$ & $0.31$\\
  \hline\hline
  \end{tabular}
    \end{table*}



\begin{table*}[htbp!]
  \centering
    \caption{Numerical values of physical parameters for the NSs  for different values of  $\zeta_6$ and $\rm D=6$}
\label{Table 3}
        \begin{tabular}{cccccccccccccccccccc}
\hline\hline  Value  & Value of Maximun & Value of  & Central & Central & & Surface & \\  
of $\zeta_6$  &  Mass $M{G_D}/(D - 3)$ & Corresponding  & Density & Pressure & $\frac{2M G_D}{ \left(D - 3\right) R^{D - 3}}$ & Redshift & \\ 
  &  in km$^{(D-3)}$  & Radius (km)  & $\widetilde{\rho}_c~(\rm MeV/{fm}^3)$ & $\widetilde{p}_c~(\rm MeV/{fm}^3)$ &  &   & 
\\ \hline 
 $8\times{10}^{14}$ & $196.57$ & $15.51$ & $1214.42$ & $288.44$ & $0.11$ & $0.06$   \\ 
$9\times{10}^{14}$ & $164.73$ & $14.72$ & $1296.14$ & $294.44$ & $0.10$ & $0.06$\\
$1\times{10}^{15}$ & $140.66$ & $13.92$ & $1476.45$ & $342.56$ & $0.10$ & $0.06$\\
$1.4\times{10}^{15}$ & $84.91$ & $11.79$ & $2039.96$ & $468.03$ & $0.10$ & $0.06$\\
$1.8\times{10}^{15}$ & $58.25$ & $10.38$ & $2653.16$ & $614.71$ & $0.10$ & $0.06$\\
  \hline\hline
  \end{tabular}
    \end{table*}


\section{Conclusion} \label{sect5}
In this study, we investigate the impact of higher dimensions on the structural characteristics and EoS of NSs. We utilize the modified Einstein field equation for D-dimensional spacetime and apply the modified DD2 model EoS to describe matter distribution within NSs, solving the fundamental equations for their structural properties. The updated DD2 model EoS in higher dimensions offers a generalized representation for the energy density ($\rho_D$) and pressure ($p_D$), enabling a comprehensive analysis of NS structure in generalized D-dimensional spacetime. Additionally, we modify and numerically solve the essential stellar structure equations for higher dimensions, systematically varying parameters such as $\rho_c G_D$, $D$, and $\zeta_D$. Our findings, depicted in various figures showcasing pressure profiles, EoS characteristics, and enclosed masses,  highlight the significant influence of higher dimensions on NS physical properties.

The analysis of mass-radius relationships across different dimensions reveals intriguing trends, showcasing the progressive increase in total mass for a constant central density as we transition to higher dimensions. The stability criterion for spherically symmetric static stellar structures, expressed by $d{M}/d{\rho_c} > 0$, holds true even in the higher-dimensional spacetime.

Furthermore, the study of radial oscillations and the assessment of stability against infinitesimal radial pulsations indicate the stability of the higher-dimensional stellar systems. The adherence to the modified Buchdahl limit for NSs is confirmed, with masses comfortably remaining within the predicted upper limit for all D-dimensional cases. The variation of surface redshift with $\tilde{M}/R$ shows a gradual decrease with increasing dimensions.

In summary, our comprehensive study contributes to a nuanced understanding of NSs in higher-dimensional spacetimes, shedding light on their structural properties, stability, and adherence to fundamental limits. These findings offer valuable insights into the behavior of dense matter in extreme conditions, furthering our knowledge of astrophysical phenomena and contributing to the broader understanding of gravity in higher dimensions.

\begin{acknowledgments}

DD and MB acknowledges the support from the Department of Atomic Energy,  Government of India through `Apex Project –Advance Research and Education in Mathematical Sciences' at  The Institute of Mathematical Sciences.  SB would like to acknowledge the financial support by DST-SERB, Govt. of India through the Core Research Grant (CRG/2020/003899) for the project entitled: `Investigating the Equation of State of Neutron Stars through Gravitational Wave Emission'.

\end{acknowledgments}




\vspace{1.0cm}

\appendix

\labelformat{section}{Appendix #1}
\labelformat{subsection}{Appendix #1}
\labelformat{subsubsection}{Appendix #1}

\section*{Appendices}

\section{The radial pulsation equations in D-dimension}\label{appnA}
Arba{\~n}il et al.~\cite{Arbanil2019} demonstrated a modified version of the Chandrasekhar radial pulsation equation~\cite{Chandrasekhar1964a,Chandrasekhar1964b}, accounting for the influence of higher dimensional spacetime. In order to assess the stability of hyperspherical objects against radial perturbations, it is necessary to examine the linearized form of the conservation of the energy-momentum tensor and the perturbed variables, including $\delta \lambda$, $\delta\nu$, $\delta \rho_D$, and $\delta p_D$, considering their temporal dependencies in the form of $e^{i\omega t}$. As a result, the radial pulsation equation for D-dimensional spacetime can be expressed as follows:
\begin{eqnarray}\label{A1.1}
&\qquad\hspace*{-1cm} \omega^2\left(p_{D}+\rho_{D}\right)e^{\lambda-\nu}\eta-\left(\frac{2}{D-3}\right)S_{D-2}G_D\left(p_{D}+\rho_{Dd}\right)\eta e^{\lambda}\,p_{Dd}\nonumber\\
&\qquad\hspace*{-1cm} +e^{-\lambda/2-\nu}\frac{d}{dr}\left(e^{\lambda/2+\nu}\frac{p_{D}\Gamma}{r^{D-2}}e^{\nu/2}\frac{d}{dr}\left(e^{-\nu/2}r^{D-2}\eta\right)\right)\nonumber\\
&\qquad\hspace*{-1cm} -\frac{2(D-2)\eta}{r}\frac{dp_{D}}{dr}+\frac{\left(p_{D}+\rho_{D}\right)\eta}{4}\left(\frac{d\nu}{dr}\right)^2=0,
\end{eqnarray}
where $\eta$ denotes Lagrangian displacement and $\Gamma~= \left\lbrace\left({\rho_D}+{p_D}\right)/{{p_D}}\right\rbrace {dp_D}/{d\rho_D}$.

It is widely recognized that Eq.~\eqref{A1.1} can be transformed into a more efficient numerical integration-friendly format. This transformation involves expressing it as a system consisting of two first-order equations, which can be represented as follows:
\begin{align}\label{A1.2}
&\quad \hspace*{-1cm} \frac{d\xi}{dr} = -\frac{1}{r}\left[(D-1){\xi} + \frac{\Delta \tilde{p}_D}{\Gamma\tilde{p}_D}\right] - \frac{d\tilde{p}_D}{dr}\frac{\xi}{\tilde{p}_D+\tilde{\rho}_D},\\ \label{A1.3}
\begin{split}
\hspace*{-0.4cm} \frac{d\Delta \tilde{p}_D}{dr} &= {\xi}\left[\omega^2 e^{\lambda-\nu}(\tilde{p}_D+\tilde{\rho}_D) r - 2 \left(D-2\right)\frac{d\tilde{p}_D}{dr}\right. \\
&\quad\hspace*{-0.4cm} \left.-\frac{2 S_{D-2} e^\lambda \tilde{p}_D\left(\tilde{\rho}_D+\tilde{p}_D\right)r}{D-3} + \left(\frac{d\tilde{p}_D}{dr}\right)^2 \frac{r}{\tilde{\rho}_D+\tilde{p}_D}\right] \\
&\quad\hspace*{-0.4cm} +\Delta\tilde{p}_D\left[\frac{1}{\tilde{\rho}_D+\tilde{p}_D}\frac{d\tilde{p}_D}{dr} - \frac{ S_{D-2} e^\lambda \tilde{p}_D\left(\tilde{\rho}_D+\tilde{p}_D\right)r}{D-3}\right],
\end{split}
\end{align}
where ${\xi}$ and $\Delta\tilde{p}_D$ denote the relative radial displacement and the Lagrangian perturbation, respectively. Note that for $D=4$ Eqs.~\eqref{A1.2} and~\eqref{A1.3} reduces to the celebrated form of the radial pulsation equations presented by Chandrasekhar~\cite{Chandrasekhar1964a,Chandrasekhar1964b}.

To numerically solve Eqs.~\eqref{A1.2} and~\eqref{A1.3}, it is essential to establish physically meaningful boundary conditions. Analogous to a vibrating string anchored at its endpoints, the radial pulsations within a star's inner region occur between its center and its surface. In light of the singularity at the origin in Eq.~\eqref{A1.2}, it is imperative that, as $r$ approaches zero, the coefficient of the $1/r$ term must vanish. This requirement yields the following boundary condition at the center for higher-dimensional compact stars:
\begin{eqnarray}\label{A1.4}
\Delta\tilde{p}_D = -(D-1)\left(\xi\Gamma\,\tilde{p}_D\right)_{\rm center}.
\end{eqnarray}
Meanwhile, at the stellar surface where $\tilde{p}_D(R) = 0$, the appropriate boundary condition is that the Lagrangian perturbation of the radial pressure must vanish:
\begin{equation}\label{A1.5}
\left(\Delta \tilde{p}_D \right)_{\rm surface}=0.
\end{equation}
On the otherhand, we consider $\xi(r=0)=1$ for normalized eigenfunction for $r\rightarrow 0$.

\section{The modified Buchdahl limit in D-dimension}\label{appnB}

To derive the equivalent of the Buchdahl's limit for D-dimensional spacetime, we differentiate Eq. ~\eqref{1.8} w.r.t $r$, and get:
\begin{eqnarray}\label{B1.1}
&\qquad\hspace{-1cm} e^{-\lambda}\left[\frac{\nu^{\prime\prime}}{r} - \frac{\nu^\prime}{r^2} - \frac{2\left(n-3\right)}{r^2} - \frac{\lambda^\prime\nu^\prime}{r} - \frac{(D-3)\lambda^\prime}{r^2}\right] \nonumber \\
&\qquad \hspace*{-1cm} + \frac{2\left(D-3\right)}{r^3}  = \frac{2}{(D-3)c^4} G_D S_{D-2} p^\prime_D.
\end{eqnarray}

Adding Eqs.~\eqref{1.6} and \eqref{1.7}, we get
\begin{eqnarray}\label{B1.2}
c^2 \rho_D + p_D = \frac{(D-3)c^4}{2G_D S_{D-2}} \frac{(\lambda^\prime+\nu^\prime)e^{-\lambda}}{r}.
\end{eqnarray}

Using the stellar structure equation $dp_D/dr = \nu^\prime/2 (\rho_D c^2 + p_D)$ and substituting Eq.~\eqref{B1.2} into Eq.~\eqref{B1.1} we find, 
\begin{eqnarray}\label{B1.3}
\qquad\hspace{-2cm} e^{-\lambda}\bigg[\frac{\nu^{\prime\prime}}{r} +  \frac{{\nu^\prime}^2}{2r} - \frac{\nu^\prime}{r^2} - \frac{\lambda^\prime\nu^\prime}{2r} - \frac{2\left(D-3\right)}{r^3} \nonumber \\ 
\qquad\hspace{1cm}  - \frac{(D-3)\lambda^\prime}{r^2}\bigg] + \frac{2\left(D-3\right)}{r^3} = 0.
\end{eqnarray}
\\
Multiplying both sides of Eq.~\eqref{B1.3} by $r$ and rearranging we get 
\begin{eqnarray}\label{B1.4}
& \qquad\hspace{-3cm} e^{-\lambda}\left[ \nu^{\prime\prime} +  \frac{{\nu^\prime}^2}{2} -\frac{\lambda^\prime\nu^\prime}{2}  + (D-3)\frac{(\nu^\prime - \lambda^\prime)}{r}\right] \nonumber \\
& \qquad\hspace{-1cm} -2\left\lbrace e^{-\lambda} \left[ \frac{\nu^\prime}{r}+\frac{(D-3)}{r^2}\right] - \frac{(D-3)}{r^2}\right\rbrace -\frac{(D-4)\nu^\prime}{r} e^{-\lambda} = 0.  \nonumber \\
\end{eqnarray}

Multiplying both sides of Eq.~\eqref{B1.4} by $2r e^{-\lambda}$ lead to the following:
\begin{eqnarray}\label{B1.5}
2r\nu^{\prime\prime} + r {\nu^\prime}^2 - r \lambda^\prime \nu ^\prime -2\nu^\prime = (D-3)\left[\frac{4}{r} \left(1-e^{\lambda}\right) + 2\lambda^\prime\right].\nonumber \\
\end{eqnarray}

Substituting the following identities, such as
\begin{eqnarray}\label{B1.6}
&\qquad\hspace{-0.8cm} \frac{d}{dr}\left[\frac{1}{r} e^{-\lambda/2}\frac{d}{dr}\left(e^{\nu/2}\right)\right] = \frac{e^{\left(\nu - \lambda\right)/2}}{4r^2}\left[2r\nu^{\prime\prime} + r{\nu^\prime}^2 -r\lambda^\prime \nu^\prime - 2\nu^\prime\right]\nonumber \hspace*{-0.9cm}  \\ \\ \label{B1.7}
\text{and}\nonumber \\
&\qquad\hspace{-3.5cm} \frac{d}{dr}\left[\frac{1-e^{-\lambda}}{2r^2}\right] = \frac{e^{-\lambda}}{2r^3}\left[r\lambda^\prime - 2\left(e^\lambda - 1\right)\right]
\end{eqnarray}

into Eq.~\eqref{B1.5} we get
\begin{eqnarray}\label{B1.7a}
\frac{c^2 e^{-\left(\nu + \lambda\right)/2}}{G_D}  \frac{d}{dr}\left[\frac{1}{r} e^{-\lambda/2}\frac{d}{dr}\left(e^{\nu/2}\right)\right] = \frac{d\bar{\rho}_{D}}{dr},
\end{eqnarray}
where we used the functional form for $e^\lambda(r)$ as given in Eq.~\eqref{2.6} and considered $\bar{\rho}_{D}$ which is defined as $\bar{\rho}_{D} = m(r)/r^{D-1}$. As $\bar{\rho}_{D}$ should monotonically decrease from the stellar central to the surface, Eq.~\eqref{B1.7a} implies
\begin{eqnarray}\label{B1.8}
\frac{d}{dr}\left[\frac{1}{r} e^{-\lambda/2}\frac{d}{dr}\left(e^{\nu/2}\right)\right] \leq 0.
\end{eqnarray}

Integrating the inequality in Eq.~\eqref{B1.8} over a radius $r$ within the star, extending to the surface with radius $R$ and multiplying both sides by $re^{\lambda/2}$, we get
\begin{eqnarray}\label{B1.9}
\frac{d e^{\nu/2}}{dr} \geq \frac{re^{\lambda/2}}{2R} e^{(\nu_0 - \lambda_0)/2} {\nu_0}^\prime.
\end{eqnarray}
Further, integrating Eq.~\eqref{B1.9} from the star's center to its surface yields
\begin{equation}\label{B2.0}
\hspace{-0.5cm} e^{\frac{\nu}{2}}\Bigg|_{r=0} \leq e^{\frac{\nu_0}{2}} - \frac{e^{\frac{\nu_0 - \lambda_0}{2}}}{2R} \nu^\prime_0 \int\limits_{0}^{r}\frac{r}{\sqrt{1 - \frac{2mG_{D}}{c^{2} (D-3) r^{D-3}}}}\,dr.
\end{equation}

As $\bar{\rho}_D$ decreases outward, we deduce $m(r)/r^{D-1} > M/r^{D-1}_0$. By replacing $m(r)/r^{D-3}$ with $\left(M/r^{D-1}\right) r^2$ and further manipulation, the modified inequality becomes
\begin{equation}\label{B2.1}
\hspace{-0.5cm} e^{\nu/2}\Bigg|_{r=0} \leq  e^{\nu_0/2} - \frac{e^\frac{\left(\nu_0 - \lambda_0\right)}{2}}{2 R}  \nu^\prime_0  \left\lbrace \frac{r^2_0}{1 - e^{-\lambda_0}} \right\rbrace \left(1- e^{-\frac{\lambda_0}{2}}\right).
\end{equation}
Considering a positive and finite pressure at the origin, $e^{\nu/2}(r=0)>0$ holds, and this leads to
\begin{equation}\label{B2.2}
e^{\nu_0/2} - \frac{e^\frac{\left(\nu_0 - \lambda_0\right)}{2}}{2 R}  \nu^\prime_0  \left\lbrace \frac{r^2_0}{1 - e^{-\lambda_0}} \right\rbrace \left(1- e^{-\frac{\lambda_0}{2}}\right) > 0.
\end{equation}

Considering Eq.~\eqref{1.8} at $r = R$, we get
\begin{eqnarray}\label{B2.3}
\frac{\nu^\prime_0}{2 R} = - \frac{\left(D -3 \right)}{2 r^2_0} + \left(D - 3\right) \frac{e^{\lambda_0}}{2 r^2_0}.
\end{eqnarray}

Substituting Eq.~\eqref{2.3} into Eq.~\eqref{2.2}, we obtain
\begin{eqnarray}\label{B2.4}
&\qquad\hspace{-2cm} e^{\nu_0/2} - \left[ - \frac{(D-3)}{2 r^2_0} + (D -3 ) \frac{e^\lambda_0}{2 r^2_0} \right] e^\frac{\left(\nu_0 - \lambda_0\right)}{2} \nonumber \\
&\qquad\hspace{1cm}  \left\lbrace \frac{r^2_0}{1 - e^{-\lambda_0}} \right\rbrace \left(1- e^{-\frac{\lambda_0}{2}}\right) > 0
\end{eqnarray}

As $e^{\nu_0/2} > 0$, $e^{\nu_0/2}$ can be removed from the inequality. Further, multiplying both sides of the equation by $e^{-\lambda_0/2}$, we get
\begin{eqnarray}\label{B2.5}
1 - e^{-\lambda_0/2} < \frac{2}{D - 1}.
\end{eqnarray}
Using Eqs.~\eqref{2.6} and~\eqref{B2.5}, we finally obtain
\begin{eqnarray}\label{B2.6}
\frac{M G_D}{c^2 \left(D - 3\right) R^{D - 3}} < \frac{2 \left(D - 2\right)}{\left(D - 1 \right)^2},
\end{eqnarray}
which represents the modified Buchdahl limit for a D-dimensional compact star with mass $M$ and total $R$. For $D = 4$ one retrieves the well known form of the Buchdahl limit~\cite{Buchdahl1959}, such as $M/R = 4/9$.


\begin{thebibliography}{90}

\bibitem{Einstein1916} A. Einstein, Sitzungsber. Preuss. Akad.Wiss. Berlin (Math. Phys.) \textbf{1916}, 688 (1916)

\bibitem{Einstein1918} A. Einstein, Sitzungsber. Preuss. Akad. Wiss. Berlin (Math. Phys.) \textbf{1918}, 154 (1918)

\bibitem{Hawking2011} S.W. Hawking and G. F. R. Ellis, The Large Scale Structure of Space-Time. Cambridge Monographs on Mathematical Physics, vol. 2 (Cambridge University Press, 2011).  

\bibitem{Wald1984} R.M. Wald, General Relativity. Chicago Univ. Pr., Chicago, USA (1984). 

\bibitem{Will2006} C.M. Will,  Living Rev. Relat. \textbf{9}, 3 (2006).  

\bibitem{Berti2015}  E. Berti et al.,  Class. Quantum Gravity \textbf{32}, 243001 (2015).  

\bibitem{Berti2018A}  E. Berti, K. Yagi,  and N. Yunes, Gen. Relat. Gravit. \textbf{50}, 46 (2018).  

\bibitem{Berti2018B}  E. Berti, K. Yagi, H. Yang, and N. Yunes,  Gen. Relat. Gravit. \textbf{50}, 49 (2018). 

\bibitem{LIGO2017}  LIGO Scientific, Virgo Collaboration, B.P. Abbott et al., Phys. Rev. Lett. \textbf{119}, 161101 (2017). 

\bibitem{LIGO2019} LIGO Scientific, Virgo Collaboration, B.P. Abbott et al., Phys. Rev. D \textbf{100}, 104036 (2019).  

\bibitem{LIGO2021}  LIGO Scientific, Virgo Collaboration, R. Abbott et al., Phys. Rev. D \textbf{103}, 122002 (2021).

\bibitem{Penrose1965} R. Penrose, Phys. Rev. Lett. \textbf{14}, 57 (1965).  
 
\bibitem{Hawking1976} S.W. Hawking, Phys. Rev. D \textbf{14}, 2460 (1976).  

\bibitem{Weinberg1989} S. Weinberg,  Rev. Mod. Phys. \textbf{61}, 1 (1989).  

\bibitem{Carroll2001} S.M. Carroll, Living Rev. Relat. \textbf{4}, 1 (2001).  

\bibitem{Padmanabhan2003} T. Padmanabhan,  Phys. Rep. \textbf{380}, 235 (2003). 

\bibitem{Birrell1984} N. Birrell and P. Davies, Quantum Fields in Curved Space. Cambridge Monographs on Mathematical Physics, vol. 2 (Cambridge University Press, Cambridge, 1984).  

\bibitem{Cardoso2018} V. Cardoso, J.A.L. Costa, K. Destounis, P. Hintz, and A. Jansen,  Phys. Rev. Lett. \textbf{120}, 031103 (2018).  

\bibitem{Rahman2019} M. Rahman, S. Chakraborty, S. SenGupta,  and A.A. Sen, J. High Energ. Phys.  \textbf{03}, 178 (2019).  

\bibitem{Abbott2017} B. P. Abbott et al.,  Astrophys. J. Lett. \textbf{848}, L12 (2017).

\bibitem{Barvinsky2003} A. O. Barvinsky and S. N. Solodukhin, Nucl. Phys. B \textbf{675}, 159 (2003).

\bibitem{Cardoso2003a} V. Cardoso, O. J. C. Dias, and J. P. S. Lemos,  Phys. Rev. D \textbf{67}, 064026 (2003).

\bibitem{Cardoso2003b} V. Cardoso, S. Yoshida, O. J. C. Dias, and J. P. S. Lemos,  Phys. Rev. D \textbf{68}, 061503 (2003).

\bibitem{Alesci2005} E. Alesci and G.Montani,  Int. J. Mod. Phys. D \textbf{14}, 923 (2005).

\bibitem{Cardoso2008} V. Cardoso, O. J. C. Dias, and P. Figueras,  Phys. Rev. D \textbf{78}, 105010 (2008).

\bibitem{Yu2017} H. Yu, B.M. Gu, F. P. Huang, Y. Q. Wang, X. H. Meng, and Y. X. Liu, J. Cosmol. Astropart. Phys. 02 (2017) 039.

\bibitem{Andriot2017} D. Andriot and G. Lucena G{\'o}mez, J. Cosmol. Astropart. Phys. \textbf{06} (2017) 048. 

\bibitem{Chakraborty2018a} S. Chakraborty, K. Chakravarti, S. Bose, and S. SenGupta, Phys. Rev. D \textbf{97}, 104053 (2018).

\bibitem{Visinelli2018} L. Visinelli, N. Bolis, and S. Vagnozzi, Phys. Rev. D \textbf{97}, 064039 (2018).

\bibitem{Pardo2018} K. Pardo, M. Fishbach, D. E. Holz, and D. N. Spergel, J. Cosmol. Astropart. Phys. 07 (2018) 048.

\bibitem{McDonough2018} E. McDonough and S. Alexander, J. Cosmol. Astropart. Phys. 11 (2018) 030.

\bibitem{Liu2023} YQ. Liu,  YQ.  Dong and YX.  Liu, Eur. Phys. J. C \textbf{83}, 857 (2023).

\bibitem{Cantiello2018} M. Cantiello, J.~B. Jensen, J.~P. Blakeslee, E. Berger, A.~J. Levan, N.~R. Tanvir, G. Raimondo, et al.. Astrophys. J. Lett. \textbf{854},  L31 (2018).

\bibitem{Ishihara2001} H. Ishihara,  Phys. Rev. Lett. \textbf{86}, 381 (2001).

\bibitem{Caldwell2001} R. Caldwell,  Phys. Lett. B \textbf{511}, 129 (2001).

\bibitem{Lin2020} Z.C. Lin, H. Yu, and Y.X. Liu,  Phys. Rev. D \textbf{101}, 104058 (2020).

\bibitem{Du2021} Y. Du, S. Tahura, D. Vaman, and K. Yagi,  Phys. Rev. D \textbf{103}, 044031 (2021).

\bibitem{Nordstrom1914a} G. Nordstr{\"o}m, Phys. Z. \textbf{15}, 504 (1914), [physics/0702221].

\bibitem{Nordstrom1914b} G. Nordstr{\"o}m, {\"O}versigt af Finska Vetenskaps-Societetens F{\"o}rhandlingar. Bd. \textbf{57}, 1 (1914), [physics/0702222].

\bibitem{Kaluza1921} T. Kaluza, Sitzungsber. Preuss. Akad. Wiss. Berlin. (Math. Phys.) \textbf{96}, 966 (1921).

\bibitem{Klein1926a} O. Klein, Zeitschrift f¨ur Physik A. \textbf{37}, 895 (1926).

\bibitem{Klein1926b} O. Klein, Nature \textbf{118}, 516 (1926).

\bibitem{Rubakov1983a} V. A. Rubakov and M. E. Shaposhnikov, Phys. Lett. B \textbf{125}, 136 (1983).

\bibitem{Rubakov1983b} V. A. Rubakov and M. E. Shaposhnikov, Phys. Lett. B \textbf{125}, 139 (1983).

\bibitem{Antoniadis1990} I. Antoniadis,  Phys.Lett. B \textbf{246}, 377 (1990).

\bibitem{Hamed1998} N. Arkani-Hamed, S. Dimopoulos, and G. Dvali,  Phys.Lett. B \textbf{429}, 263 (1998).

\bibitem{Antoniadis1998} I. Antoniadis, N. Arkani-Hamed, S. Dimopoulos, and G. Dvali,  Phys.Lett. B \textbf{436},  257 (1998).

\bibitem{Csaki2004} C. Csaki,  ``TASI lectures on extra dimensions and branes," in From fields to strings: Circumnavigating theoretical physics.  Ian Kogan memorial collection (3 volume set), pp. 605 (2004). [ arXiv:hep-ph/0404096]

\bibitem{Perez2005} A. Perez-Lorenzana,  J. Phys. Conf. Ser.  \textbf{18}, 224 (2005).

\bibitem{Sundrum2005} R. Sundrum,  ``Tasi 2004 lectures: To the fifth dimension and back," in Theoretical Advanced Study Institute in Elementary Particle Physics: Many Dimensions of String Theory (TASI 2005) Boulder, Colorado, June 5-July 1, 2005, pp. 585 (2005).  [arXiv:hep-th/0508134]

\bibitem{Dadhich2000} N. Dadhich, R. Maartens, P. Papadopoulos, V. Rezania, Phys. Lett. B \textbf{487}, 1 (2000).

\bibitem{Chamblin2000} A. Chamblin, S.W. Hawking, H.S. Reall, Phys. Rev. D \textbf{61}, 065007 (2000). 

\bibitem{Emparan2000} R. Emparan, G.T. Horowitz, R.C. Myers,  JHEP \textbf{01}, 007 (2000). 

\bibitem{Chamblin2001} A. Chamblin, H.S. Reall, H.-A. Shinkai, T. Shiromizu, Phys. Rev. D \textbf{63}, 064015 (2001).  

\bibitem{Harko2004} T. Harko and M.K. Mak, Vacuum solutions of the gravitational field equations in the brane world model. Phys. Rev. D \textbf{69}, 064020 (2004).  

\bibitem{Aliev2005} A.N. Aliev and A.E. Gumrukcuoglu, Phys. Rev. D \textbf{71}, 104027 (2005).  

\bibitem{Chakraborty2016} S. Chakraborty and S. SenGupta, Class. Quantum Gravity \textbf{33},  225001 (2016).  

\bibitem{Nakas2021} T. Nakas and P. Kanti, Phys. Lett. B \textbf{816}, 136278 (2021).  

\bibitem{Csaki1999} C. Csaki, M. Graesser,  C.F. Kolda,  and J. Terning, Phys. Lett. B \textbf{462}, 34 (1999).  

\bibitem{Csaki2000} C. Csaki, M. Graesser, L. Randall,  and J. Terning,  Phys. Rev. D \textbf{62}, 045015 (2000). 

\bibitem{Harko1993} T. Harko,  Acta Physica Hungarica \textbf{73}, 165 (1993).

\bibitem{Harko2000} T. Harko and M. Mak, J. Math. Phys. (N.Y.) \textbf{41}, 4752 (2000).

\bibitem{Leon2000} J. Ponce de Leon and N. Cruz,  Gen. Relativ. Gravit. \textbf{32}, 1207 (2000).

\bibitem{Paul2001} B. C. Paul, Class. Quantum Gravity \textbf{18}, 2637 (2001).

\bibitem{Dadhich2017} N. Dadhich and S. Chakraborty,  Phys. Rev. D \textbf{95}, 064059 (2017).

\bibitem{Chakraborty2018} S. Chakraborty and S. Sengupta, J. Cosmol. Astropart. Phys. 05 (2018) 032.

 \bibitem{Ghosh2000} S. G. Ghosh and A. Beesham,  Classical Quantum Gravity \textbf{17}, 4959 (2000).

\bibitem{Ghosh2001a}  S. G. Ghosh and A. Beesham,  Phys. Rev. D \textbf{64}, 124005 (2001).

 \bibitem{Ghosh2001b} S. G. Ghosh and N. Dadhich,  Phys. Rev. D \textbf{64}, 047501 (2001).

\bibitem{Chavanis2017} P.-H. Chavanis, Phys. Rev. D \textbf{76}, 023004 (2017).

\bibitem{Arbanil2019} J.~D.~V.  Arba{\~n}il, G.~A.  Carvalho,  R.~V.  Lobato, R.~M.  Marinho, and M.  Malheiro, Phys. Rev.  D \textbf{100}, 024035 (2019).

\bibitem{Arbanil2020} Jos{\'e} D.~V. Arba{\~n}il, C{\'e}sar H. Lenzi, and Manuel Malheiro, Phys. Rev. D \textbf{102}, 084014 (2020).

\bibitem{Hempel2010}M. Hempel and J. Schaffner-Bielich, Nucl. Phys. A \textbf{837}, 210 (2010). 

\bibitem{Banik2014} S. Banik, M. Hempel, and D. Bandyopadhyay, Astrophys. J. Suppl. Ser. \textbf{214}, 22 (2014). 

\bibitem{Banik2021} T. Malik, S. Banik, and D. Bandyopadhyay, Astrophys. J. \textbf{910}, 96 (2021).

\bibitem{Tolman1939} R. C. Tolman, Phys. Rev. D \textbf{55}, 364 (1939).

\bibitem{Oppenheimer1939} J. R. Oppenheimer and G. Volkoff,  Phys. Rev. D \textbf{55}, 374 (1939).

\bibitem{Typel2005}  S. Typel, Phys. Rev. C \textbf{71}, 064301 (2005).

\bibitem{Typel2010} S. Typel, G. R{\"o}pke, T. Kl{\"a}hn, D. Blaschke, and H. H. Wolter,  Phys. Rev. C \textbf{81}, 015803 (2010).

\bibitem{Typel1999} S. Typel and H. H. Wolter, Nucl. Phys. A \textbf{656}, 331 (1999).

\bibitem{Buchdahl1959} H. A. Buchdahl,  Phys. Rev. \textbf{116}, 1027 (1959).

\bibitem{Chandrasekhar1964a} S. Chandrasekhar,  Astrophys. J. \textbf{140}, 417 (1964).

\bibitem{Chandrasekhar1964b} S. Chandrasekhar,  Phys. Rev. Lett. \textbf{12}, 114 (1964).

\end{thebibliography}
\end{document}